\documentclass{jpp}
\usepackage{graphicx}
\usepackage{epstopdf, epsfig}
\usepackage{hyperref}
\usepackage{natbib}
\usepackage{amsmath}
\usepackage{float}

\renewcommand{\vec}{\boldsymbol}
\newcommand{\Del}{\vec{\nabla}}

\shorttitle{Rotating MHD turbulence}
\shortauthor{N. K. Bell and S. V. Nazarenko}

\title{ Rotating magnetohydrodynamic turbulence}

\author{N. K. Bell\aff{1}
 \and S. V.  Nazarenko\aff{2}  \corresp{\email{Sergey.Nazarenko@unice.fr}}
}

\affiliation{\aff{1}Mathematical Institute, University of Warwick,
Coventry, CV4 7AL, UK
\aff{2}L'Institut de Physique  de Nice, Universit\'{e} Nice-Sophia Antipolis, Parc Valrose, 06108 Nice, France}

\begin{document}

\maketitle

\begin{abstract}
Turbulence in rotating Magneto-hydrodynamic systems is studied theoretically and numerically.
In the linear limit, when the velocity and magnetic perturbations are small, the system supports two types
of waves. When the rotation effects are stronger than the ones of the external magnetic field, one of these
waves contains most of the kinetic energy (inertial wave) and the other -- most of the magnetic energy
(magnetostrophic wave).
The weak wave turbulence (WWT) theory for decoupled inertial and magnetospheric wave systems was previously
derived by~\citet{Galtier2014b}. In the present paper, we derive theory of strong turbulence for such waves based on the critical balance (CB) approach conjecturing that the linear and nonlinear timescales are of similar magnitudes in a wide range of turbulent scales.
Regimes of weak and strong wave turbulence are simulated numerically. The results appear to be in good agreement with the WWT and CB predictions, particularly for the exponents of the kinetic and magnetic energy spectra.
\end{abstract}

\section{Introduction}
Astrophysical flows are generally electrically conducting prompting the use of Magneto-Hydrodynamics (MHD) for describing their dynamics. Such flows are often accompanied by large-scale magnetic fields which results in the existence of a rich zoo of  wave modes. This has motivated the use of weak wave turbulence (WWT) theory when studying MHD systems~\citep{Galtier2000, Galtier2005, Galtier2006, Nazarenko2007, Schekochihin2012}. WWT theory describes the long-time statistical behaviour  of weakly nonlinear dispersive waves~\citep{ZakharovBook, NazarenkoBook}.  In most physical situations however, there is a coexistence of strongly nonlinear coherent structures and weakly nonlinear waves. WWT theory is applicable when the timescale of the linear waves is much shorter than the nonlinear timescale. In strong wave turbulence, there may exist a regime where these two timescales are of the same order over a wide range of scales. This is the so-called critical balance (CB) conjecture which was introduced in the context of MHD by~\citet{Goldreich1995b}. 

Along with the presence of a mean magnetic field, astrophysical flows often undergo rotation about an axis. Rotating MHD then has wide application including planetary flows, stellar flows and accretion discs. 
An incompressible MHD system under solid body rotation and in the presence of a uniform background magnetic field will be considered in this paper. The governing equations in the rotating frame of reference are:
\begin{equation} \label{Momentum}
\frac{\partial \vec{u}}{\partial t} + 2\vec{\Omega_0} \times \vec{u} + \vec{u}\cdot\vec{\nabla}\vec{u} = -\Del P_\ast + \vec{b_0} \cdot\Del\vec{b} + \vec{b}\cdot \Del \vec{b} + \nu\Del^2\vec{u},
\end{equation} 
\begin{equation}
\frac{\partial\vec{b}}{\partial t} + \vec{u}\cdot\Del\vec{b}=\vec{b_0}\cdot\Del\vec{u} + \vec{b}\cdot\Del\vec{u}+\eta\Del^2\vec{b},
\end{equation}
\begin{equation}
\Del\cdot\vec{u}=0,
\end{equation}
\begin{equation} \label{incompressiblecondition}
\Del\cdot\vec{b}=0,
\end{equation}
where $\vec{v}$ is the velocity, $P_\ast$ is the total pressure, $\vec{b}$ is the magnetic field normalised to a velocity, $\vec{b_0}$ is the uniform normalised magnetic field, $\Omega_0$ is the rotation rate and $\nu$ and $\eta$ are the kinematic viscosity and magnetic diffusivity respectively. For the remainder of this paper we shall assume that the axis of rotation is aligned with the background magnetic field,
\begin{equation}
\vec{\Omega_0}=\Omega_0\vec{\hat{e}_\parallel}, \qquad \vec{b_0}=b_0 \vec{\hat{e}_\parallel},
\end{equation}
where $\vec{\hat{e}_\parallel}$ is a unit vector.

The addition of the Coriolis force from rotation yields   dynamical  effects the importance of which is measured by the Rossby number,
\begin{equation}
Ro = \frac{U_0}{L_0\Omega_0},
\end{equation}
where $U_0$, $L_0$ and $\Omega_0$ are typical velocity, length scale and rotation rate respectively. The Rossby number is the ratio of the advection term and the Coriolis force in the Navier-Stokes equations. Small Rossby numbers thus correspond to flows in which the rotation is of significant importance as is the case for planetary flows especially in the case of large planets~\citep{Shirley1997,Roberts2013}. 

In rotating MHD there is also another important parameter -- the magneto-inertial length,
\begin{equation}
d=\frac{b_0}{\Omega_0}.
\end{equation} 
The value of $d$ determines the relative strength of  the Lorentz force and the Coriolis force. We find for the ratio of these two forces:
\begin{equation}
\mathcal{D} = \frac{|\vec{b_0}\cdot\Del\vec{b}|}{|2\vec{\Omega_0}\times\vec{u}|} \sim \frac{b}{u}\frac{d}{l}.
\end{equation}
When $\mathcal{D}$ is large, the Lorentz force is dominant and when it is  small the Coriolis force is dominant. Assuming that $u$ and $b$ are known, the size of $\mathcal{D}$ can be determined using $d/l$, or equivalently, $kd$ in Fourier space.

There have been various numerical studies of the effects of rotation in hydrodynamic turbulence, for example~\citep{Mininni2010a, Mininni2012, Teitelbaum2009}. The theoretical groundwork for WWT regime of the rotating turbulence has been developed by~\citet{Galtier2003}.  A prediction for the energy spectrum $E(k)\sim k_\perp^{-5/2}$ was found which applies when the  waves in the rotating fluid (inertial waves) are weakly nonlinear.  A prediction for strong  turbulence based on CB was found by~\citet{Schekochihin2012} -- it gives the energy spectrum $E(k)\sim k_\perp^{-5/3}$. More recently a WWT theory has been developed for rotating MHD~\citep{Galtier2014b}. Predictions were made for the energy spectra in two asymptotic regions: $kd=k b_0/\Omega_0 \rightarrow \infty$ and $kd\rightarrow 0$. As $kd\rightarrow\infty$, the linear waves collapse onto the Alfv\'{e}n waves resulting in a $k^{-2}$ spectrum as first found in~\citet{Galtier2000}. In the $kd\rightarrow 0 $ limit, the left polarised waves (inertial waves) and the right polarised waves (magnetostrophic waves) become separate such that  the inertial waves contain most of the kinetic energy and the magnetostrophic waves contain most of the magnetic energy. Thus, when the two types of waves are decoupled, the kinetic energy spectrum is therefore the same as for rotating hydrodynamic turbulence. Incidentially, the magnetic energy spectrum was also found to be $E(k)\sim k_\perp^{-5/2}$. 

In this paper, we present numerical simulations of rotating MHD turbulence. We begin with an overview of the WWT theory and extend it to make a prediction of the energy spectra based upon the CB phenomenology. We then present two numerical simulations which aim to test the theoretical predictions. 

\section{Wave modes in rotating MHD}

Linear waves in rotating MHD are circularly polarized and dispersive. The general solution for the frequency is given by
\begin{equation}
\omega\equiv\omega_\Lambda^s = \frac{sk_\parallel\Omega_0}{k}\left( -s\Lambda + \sqrt{1+k^2 d^2}\right),
\end{equation}
where $s=\pm 1$ defines the directional polarity such that we always have $sk_\parallel>0$ and $\Lambda=\pm s$ gives the circular polarity with $\Lambda=s$ indicating right polarization and $\Lambda=-s$ left polarization. The linear waves can be considered in the limits $kd\rightarrow \infty$ and $kd\rightarrow 0$. In the small scale limit ($kd\rightarrow 0$), the frequency of the right and left polarized waves collapse onto the Alfv\'{e}n wave frequency. In this limit the turbulence properties can therefore be studied as Alfv\'{e}n wave turbulence for which there exists a large body of work, for example~\citep{Goldreich1995b, Galtier2000, Galtier2001, Schekochihin2012, Meyrand2015}. In the large scale limit ($kd\rightarrow 0$), the right and left polarized waves are the pure magnetostrophic waves and the pure inertial waves respectively. The frequencies are given by
\begin{equation}
\omega_M\equiv \omega_s^s = \frac{sk_\parallel k d b_0}{2},
\end{equation}
\begin{equation}
\omega_I\equiv \omega_{-s}^s = \frac{2s\Omega_0 k_\parallel}{k}.
\end{equation}
For the magnetostrophic branch one can typically assume a balance between the Coriolis and Lorentz forces~\citep{Finlay2008}. Following this assumption, the nonlinear evolution of the magnetic field under relatively strong rotation and uniform magnetic field can be described by the magnetostrophic equation \citep{Galtier2014b}
\begin{equation} \label{MagnetostrophicEq}
\frac{\partial\vec{b}}{\partial t} = -\frac{d}{2}\Del\times\left[ (\Del\times\vec{b}) \times (\vec{b}+\vec{b_0})\right] +\eta\Del^2\vec{b}.
\end{equation}
This equation will be used to study the wave turbulence properties for the magnetostrophic waves. Note, however, that the balance between the Coriolis and Lorentz forces may
break at the level of the nonlinear terms if the scales of the inertial and magnetostrophic waves are separated. In this case the two types of waves get coupled (see the second part of the appendix).

\section{Weak wave turbulence}
\subsection{Energy cascade spectra}
The WWT theory for rotating MHD was developed by~\citet{Galtier2014b}. The derivation will not be reproduced here but the results which provide context to our numerical simulations will be discussed. We turn our attention to the large scale limit $kd\rightarrow 0$. In this limit and assuming that nonlinear interactions occur locally in $k$-space it can be shown that in the leading order the inertial waves decouple from the magnetostrophic waves. This is because no resonance triads simultaneously containing both types of waves exist in this regime. A discussion of this decoupling and identification of a coupled regime can be found in the appendix to this paper.  Furthermore in this limit, the inertial waves contain all of the kinetic energy and the magnetostrophic waves contain all of the magnetic energy. Both the inertial wave turbulence and the magnetostrophic wave turbulence are found to become anisotropic such that $k_\perp\gg k_\parallel$. 

Let us first consider the inertial waves.
The WWT theory applies when the dynamics are dominated by the weakly nonlinear waves. In terms of time scales this implies that the period of the linear waves is much shorter than the nonlinear turnover time,
\begin{equation}
\tau_I \ll \tau_{nl}.
\end{equation}
Looking first at the momentum equation~(\ref{Momentum}), the nonlinear time scale is
\begin{equation} \label{eq:momnl}
\tau_{nl} \sim \frac{1}{k\widetilde{U}}
\end{equation}
where $\widetilde{U}$ is the oscillating velocity, and the period of the inertial waves is 
\begin{equation} \label{eq:inertialtime}
\tau_{I} \sim (\omega_I)^{-1} = \frac{k}{2s\Omega_0 k_\parallel}.
\end{equation}
For a phenomenological derivation of the energy spectrum, we require the characteristic transfer time. If we assume that after a number of stochastic wavepacket collisions the cumulative effect may be regarded as a random walk, then we can use~\citep{Iroshnikov1964, Kraichnan1965}
\begin{equation}
\tau_{tr} \sim \frac{\tau_{nl}^2}{\tau_I}.
\end{equation}
Assuming a stationary state in which the kinetic energy flux per unit mass ($\epsilon^u$) is independent of scale we find
\begin{equation}
\epsilon^u \sim \frac{E^u}{\tau_{tr}} 
\sim \frac{E^u(k_\perp,k_\parallel)k_\perp k_\parallel}{\tau_{tr}}.
\end{equation}
Then making use of $\widetilde{U}^2\sim E(k_\perp,k_\parallel)k_\perp k_\parallel$ and the anisotropic assumption $k_\perp\gg k_\parallel$ we have, after some algebra ~\citep{Galtier2003},
\begin{equation} \label{WTinertialspec}
E^u(k_\perp,k_\parallel)\sim \sqrt{\epsilon^u \Omega_0} k_\perp^{-5/2}k_\parallel^{-1/2}.
\end{equation}

To study the magnetostrophic waves, we look at the magnetostrophic equation~(\ref{MagnetostrophicEq}). The nonlinear time scale is 
\begin{equation} \label{eq:magnl}
\tau_{nl} \sim \frac{1}{k_\perp^2 d \widetilde{B}}
\end{equation}
with $\widetilde{B}$ being the oscillating magnetic field,
and the period of magnetostrophic waves is given by
\begin{equation} \label{eq:magtime}
\tau_M \sim (\omega_M)^{-1} = \frac{2}{sk_\parallel k d b_0}.
\end{equation}
Performing the same analysis as for the kinetic energy spectrum, we find the magnetic energy spectrum ~\citep{Galtier2014b}:
\begin{equation} \label{WTmagspec}
E^{b}(k_\perp,k_\parallel)\sim \sqrt{\frac{\epsilon^b b_0}{d}}k_\perp^{-5/2}| k_\parallel |^{-1/2}.
\end{equation}

The same phenomenology can be applied to the other inviscid invariant of rotating MHD, the hybrid helicity ~\citep{Galtier2014b}. 
Following the dual cascade argument, one can conclude that the hybrid helicity should cascade toward smaller $k$'s (inverse cascade) whereas the energy -- toward larger $k$'s (direct cascade).
  The hybrid helicity cascade is an interesting topic for future study with possible insight into a dynamo process. 
  In the present paper, we consider the direct energy cascade only.

\subsection{Domain of validity for weak wave turbulence approach}
WWT theory relies upon the time scale separation between the linear time scale (wave period) and the nonlinear time scale, i.e. 
\begin{equation}
\frac{\tau_l}{\tau_{nl}}\ll 1.
\end{equation}
In rotating MHD, there are two linear wave branches with  different dispersion relations. We have defined different nonlinear times for each of these branches, given by Equation~(\ref{eq:momnl}) for the inertial waves and Equation~(\ref{eq:magnl}) for the magnetostrophic waves. For the respective ratios of these times we have:
\begin{equation}
\label{r1}
\chi^u=\frac{\tau_{I}}{\tau^u_{nl}}\sim \frac{k_\perp^2 \widetilde{U}}{k_\parallel \Omega_0} ,
\end{equation}
\begin{equation}
\label{r2}
\chi^b=\frac{\tau_{M}}{\tau^b_{nl}}\sim \frac{k_\perp \widetilde{B}}{k_\parallel b_0} .
\end{equation}
For the WWT theory to apply, the following two relations must hold: $\chi^u \ll 1$ and $\chi^b \ll 1$.
It is clear  that the ratios will vary with the wavenumber and thus may not remain uniformly small across the entire wavenumber range. The dependence of both $\chi^u$ and $\chi^b$ can be estimated using the WWT predictions along with $\widetilde{U}\sim \sqrt{E^u(k_\perp,k_\parallel)k_\perp k_\parallel}$ and $\widetilde{B}\sim \sqrt{E^b(k_\perp,k_\parallel)k_\perp k_\parallel}$. One then finds
\begin{equation}
\chi^u\sim \left( \frac{\sqrt{\epsilon^u}}{\Omega_0}\right)^{1/2}k_\parallel^{-3/4}k_\perp^{5/4},
\end{equation}
\begin{equation}
\chi^b\sim \left(\frac{\sqrt{\epsilon^b}}{b_0 d}\right)^{1/2}k_\parallel^{-3/4}k_\perp^{1/4}.
\end{equation}
Both $\chi^u$ and $\chi^b$ grow as $k_\perp$ increases and so there will be some scale (provided that dissipation is weak enough) at which the WWT assumption is broken and where the CB assumption becomes relevant. Such a transition from weak to strong wave turbulence has been observed in numerical simulations of Alfv\'{e}n wave turbulence~\citep{Meyrand2016} and Hall MHD turbulence~\citep{Meyrand2017}. 

\subsection{Strong wave turbulence}
In strong wave turbulence, it is natural to assume that the energy spectrum saturates when the nonlinear interaction time becomes of the same order as the linear wave period over a wide range of turbulent scales~\citep{NazarenkoBook}. Such states are known as a CB as introduced in MHD turbulence by~\citet{Goldreich1995b}. It has been proposed that CB provides a universal scaling conjecture for determining the spectra of strong turbulence in anisotropic wave systems~\citep{Nazarenko2011}. 

In the classical Kolmogorov spectrum of turbulence~\citep{Kolmogorov1941a}, the system is isotropic and highly nonlinear such that the transfer time is simply the nonlinear timescale. When waves are present there is an additional time scale, the period of the linear waves. The CB assumption
\begin{equation} \label{critbalance}
\tau_{tr}\sim\tau_l\sim \tau_{nl}
\end{equation}
provides the additional scaling required to perform a heuristic derivation of the energy spectrum. 

To derive the kinetic energy spectrum in the CB regime we shall again turn our attention to the momentum equation. The nonlinear time scale and the linear wave period are again given by~(\ref{eq:momnl}) and~(\ref{eq:inertialtime}) respectively. Balancing these time scales as per~(\ref{critbalance}) gives
\begin{equation} \label{eq:critbalinertialU}
\widetilde{U}\sim \frac{\Omega_0 k_\parallel}{k_\perp^2}.
\end{equation}
The kinetic energy flux is then found to be
\begin{equation}
\epsilon^u\sim \frac{\widetilde{U}^2}{\tau_{tr}} 
\sim \widetilde{U}^3 k_\perp
\sim \frac{\Omega_0^3 k_\parallel^3}{k_\perp^5},
\end{equation}
which, when rearranged, gives the wavenumber scaling for the CB ~\citep{Nazarenko2011},
\begin{equation} \label{eq:critbalinertialwave}
k_\parallel\sim (\epsilon^u)^{1/3}\Omega_0^{-1}k_\perp^{5/3}. 
\end{equation}
Using relations~(\ref{eq:critbalinertialU}) and~(\ref{eq:critbalinertialwave}) we can derive the kinetic energy spectrum,
\begin{equation} \label{CBinertialspec}
E^u(k_\perp)\sim \frac{\widetilde{U}^2}{k_\perp}
\sim (\epsilon^u)^{2/3}k_\perp^{-5/3}.
\end{equation}

The magnetic energy spectrum can be calculated similarly using the magnetostrophic equation in which the nonlinear time scale and the linear wave period are given by~(\ref{eq:magnl}) and~(\ref{eq:magtime}) respectively. Equating these two time scales gives a scaling for the wave amplitude as 
\begin{equation}
\widetilde{B}\sim \frac{k_\parallel b_0}{k_\perp}
\end{equation}
and the magnetic energy flux as
\begin{equation}
\epsilon^b\sim \frac{\widetilde{B}^2}{\tau_{tr}}.
\end{equation}
The CB scaling for the wavenumbers is thus
\begin{equation}
k_\parallel\sim (\epsilon^b)^{1/3}b_0^{-1} d^{-1/3}k_\perp^{1/3},
\label{cbm1}
\end{equation}
which leads to the magnetic energy spectrum
\begin{equation} \label{CBmagspec}
E^b(k_\perp)\sim \left(\frac{\epsilon^b}{d}\right)^{2/3}k_\perp^{-7/3}.
\end{equation}
Expressions (\ref{cbm1}) and (\ref{CBmagspec}) are the new theoretical predictions for the strong wave turbulence of magnetostrophic waves
which will be put to test (together with the previous predictions) by numerical simulations in the present paper.

\section{Numerical simulations}
\subsection{Set-up}
The rotating MHD equations~(\ref{Momentum})--(\ref{incompressiblecondition}) were solved numerically using the Fourier pseudospectral code GHOST~\citep{Gomez2005,Mininni2007,Mininni2011}. Time integration is performed by a second-order Runge-Kutta scheme and the $2/3$ rule is employed for dealiasing. An isotropic initial condition of velocity and magnetic field fluctuations with random phases was chosen. We consider a decaying turbulence to avoid any artefacts from  forcing. A hyperviscosity $\nu \Del^6 \vec{u}$ and hyperdiffusivity $\eta \Del^6 \vec{b}$ were used in place of the viscous and diffusive terms.

In order to access the two regimes of rotating MHD turbulence, namely the WWT and the CB regimes,  we will be interested in the large-scale limit,
\begin{equation}
kd = k\frac{b_0}{\Omega_0} \ll 1,
\end{equation}
and the limit of small Rossby numbers,
\begin{equation}
Ro \sim \frac{k U}{\Omega_0}\ll 1.
\end{equation}
Again, these conditions may be well satisfied at one end of the wavenumber range (small $k$'s) and only marginally at the other (large $k$'s).
The difference between the weak and strong turbulence regimes is controlled by the ratios of linear wave period to nonlinear turnover time given by equations~(\ref{r1}) and (\ref{r2}). These ratios should be very small in the WTT limit and of the order $1$ in the CB. 

\subsection{Simulation $A$: critical balance}
In simulation $A$ we have aimed to access the CB regime. The parameters used for the simulation can be found in table~\ref{tab:para}. An isotropic initial condition was chosen in the range $k=\left[2,4\right]$. The simulation has been performed in a periodic box of spatial resolution $512^3$. After application of the $N/3$ rule for dealiasing, this gives a maximum wavenumber of $170$. The parameter $kd$ therefore ranges from $0$ to $8.5$. This is only moderately small for about a decade of the wavenumber range and thus complete decoupling of the magnetic and kinetic energies may not be achieved. However, it is sufficiently far from the opposite limit $kd\rightarrow \infty$ for this simulation, so the  inertial and the magnetostrophic wave branches are well separated. The strength of the initial condition is such that the ratios of linear  to nonlinear time scales are of the order unity as is expected for a CB state. 

\begin{table}
  \begin{center}
\def~{\hphantom{0}}
  \begin{tabular}{lcccccccccc}
      Simulation  & $b_0$   &   $\Omega_0$ & $\widetilde{U}$ & $\widetilde{B}$ & $\nu$ & $\eta$  \\[3pt]
       A   & 50 & 1000 & 5.0 & 15.0 & $10^{-12}$ & $10^{-12}$ \\
       B  & 50 & 1000 & 0.5 & 1.5 & $10^{-14}$ & $10^{-14}$ 
  \end{tabular}
  \caption{Parameters used in the numerical simulations.}
  \label{tab:para}
  \end{center}
\end{table}

\begin{figure}
  \centerline{\includegraphics[height=5cm, width=14cm]{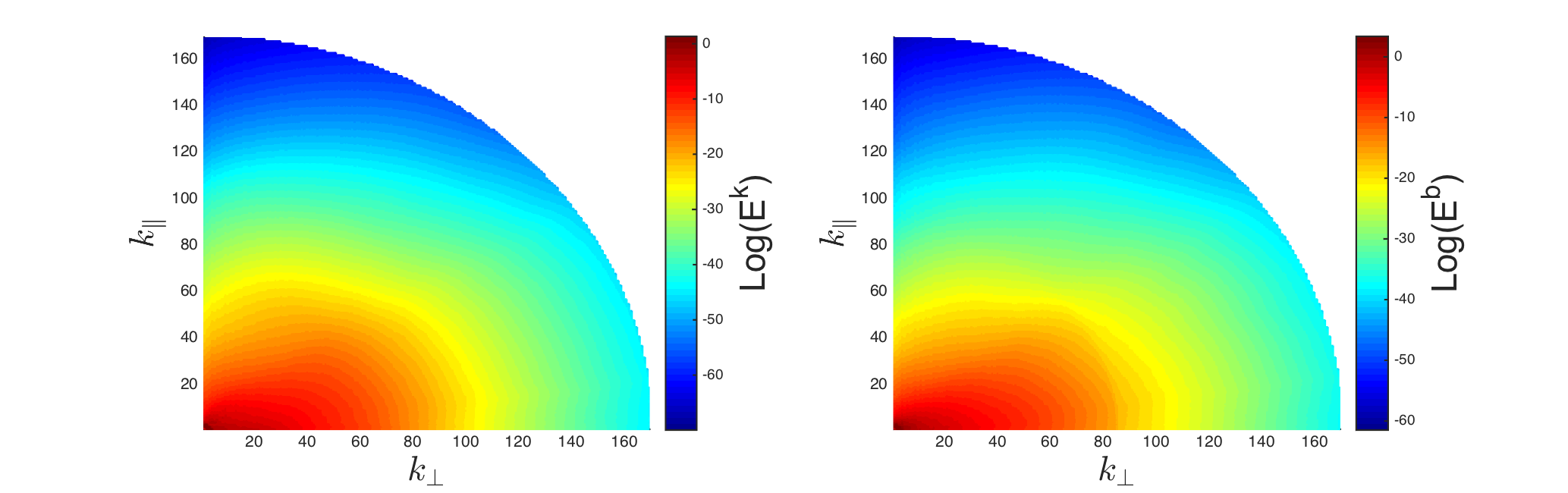}}
  \caption{$2D$ energy spectra for the (\textit{a}) kinetic\protect ~ and (\textit{b}) magnetic energies.}
\label{fig:CBAni}
\end{figure}

\begin{figure}
  \centerline{\includegraphics[width=1.1\textwidth]{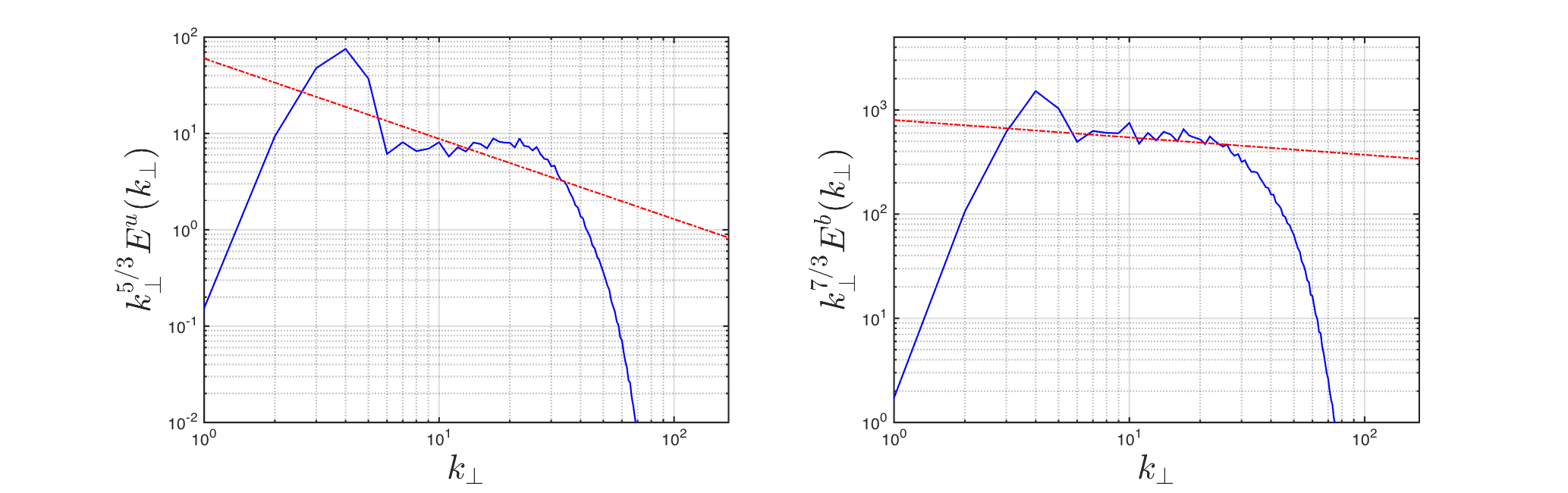}}
  \caption{Compensated axisymmetric energy spectra for the (\textit{a}) kinetic\protect ~ and (\textit{b}) magnetic energies. The straight lines indicate the WWT prediction for comparison.}
\label{fig:CBaxispec}
\end{figure}

The $2D$ energy spectra are plotted in Figure~\ref{fig:CBAni}. For both the kinetic and magnetic energy there is a preferential transfer of energy along $k_\perp$ leading to the anisotropy $k_\perp > k_\parallel$. Figure~\ref{fig:CBaxispec} shows the axially averaged energy spectra integrated over $k_\parallel$. Each spectrum is compensated by the relevant CB prediction and is plotted alongside the WWT slope prediction for comparison. Both the kinetic and the magnetic energy spectra show an excellent agreement with the CB predictions. The WWT prediction for the kinetic energy is clearly far from the observed spectrum. For the magnetic energy spectrum the WWT and the CB predictions are close, and they fit the observed spectrum equally well.

So far we have concentrated on the spatial properties of rotating MHD at a given point in time. In order to identify the presence of waves however, it is necessary to consider the properties in the spatio-temporal domain. A common method is to analyse the spatio-temporal spectrum~\citep{NAZARENKO20061,Nazarenko2007a,PhysRevB.84.064516,Leoni2015}. This can then be compared with the dispersion relation of the linear waves. Computing the spatio-temporal spectrum requires simultaneous space and time Fourier transforms. In order to resolve all the waves, the time sampling frequency must be at least twice as large as the frequency of the fastest waves in the system and the total acquisition time must be larger than both the slowest wave period and the turnover time of the slowest eddies. These requirements result in a high storage space requirement which proved restrictive, so a subset of data was collected for a fixed $k_\parallel=3$. The minimum and maximum time periods for the inertial and magnetostrophic waves in our system are
\begin{equation}
\begin{aligned}
(\tau_I)_{min}\sim 0.00016 \qquad (\tau_I)_{max}\sim 0.013 \\
(\tau_M)_{min}\sim 0.003 \qquad (\tau_M)_{max}\sim 0.26.
\end{aligned}
\end{equation}
\begin{figure}
  \centerline{\includegraphics[width=0.75\textwidth]{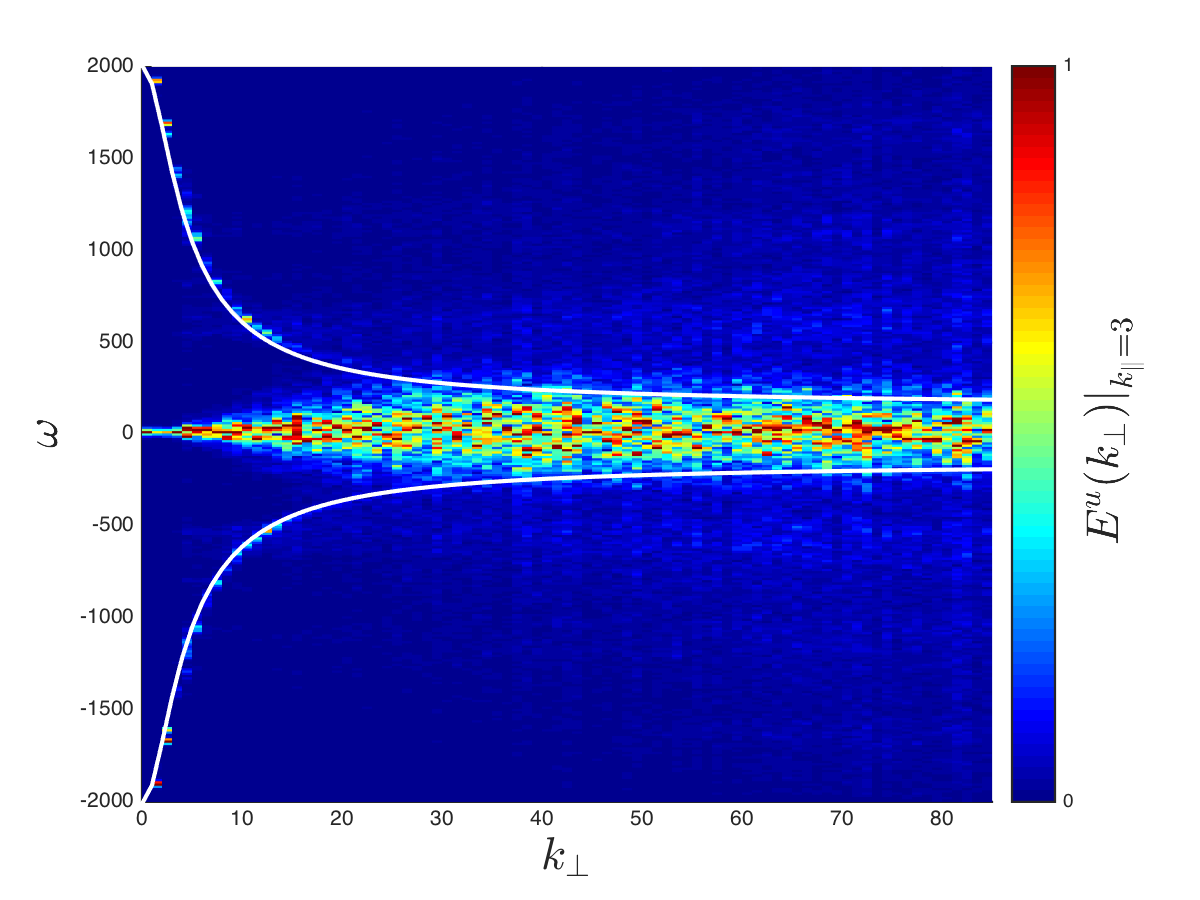}}
  \caption{Spatio-temporal energy kinetic energy spectrum in simulation A. The white lines indicate the dispersion relation for inertial waves.}
\label{fig:CBkinfreq}
\end{figure}

\begin{figure}
  \centerline{\includegraphics[width=0.75\textwidth]{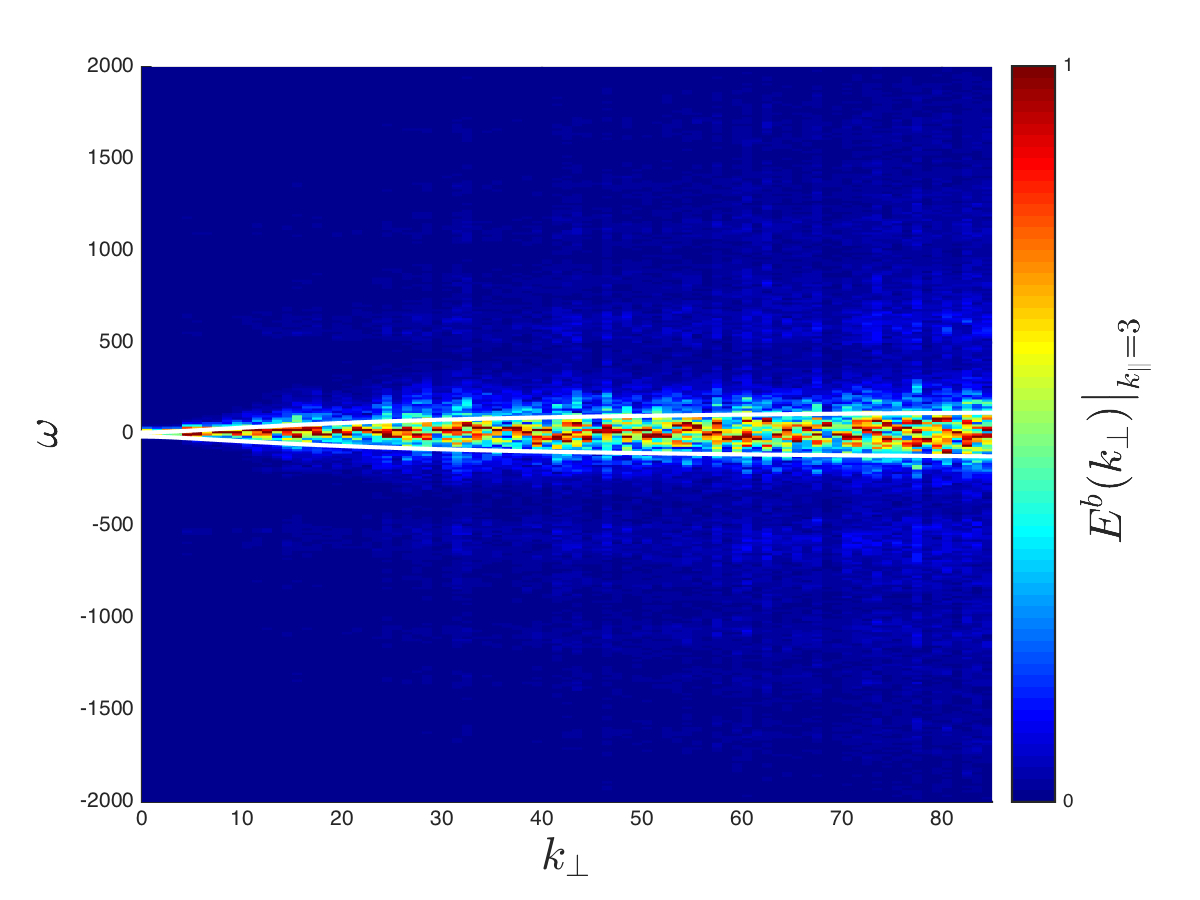}}
  \caption{Spatio-temporal energy magnetic energy spectrum in simulation A. The white lines indicate the dispersion relation for magnetostrophic waves.}
\label{fig:CBmagfreq}
\end{figure}
We used an acquisition frequency of $\delta t =0.001$ and $T = \delta t = 0.511$. This resolves all of the magnetostrophic waves and the inertial waves for $k_\perp > 6$. In Figures~\ref{fig:CBkinfreq} and~\ref{fig:CBmagfreq} we plot the spatio-temporal spectra for the kinetic and magnetostrophic energy respectively. The spectra are calculated at a fixed $k_\parallel=3$ and overlayed with the inertial and magnetostrophic wave dispersion relations. In the WWT regime, one would expect the kinetic and magnetic energy to accumulate narrowly on and near the dispersion relations for the linear waves. In a CB regime however,  the width of the distribution in the $\omega$-direction is of the same size as the linear wave frequency. It was noted before for the Alfven wave turbulence that in the CB regime  the dispersion relation acts as a boundary with the energy filling the region below it~\citep{TenBarge2012}. In Figure~\ref{fig:CBmagfreq} we see that the magnetic energy behaves exactly like this with the energy filling the area in $(k_\perp,\omega)$ space between the two polarities of the magnetostrophic dispersion relation. In Figure~\ref{fig:CBkinfreq} we see that the kinetic energy behaves in a similar way at high wavenumbers but does not fill the region at small wavenumbers. The ratio of linear wave frequency to the inverse nonlinear turnover time becomes less than $1$ at 
$k_\perp<24$ 
and we can clearly see a separation of two branches--weak waves and a low-frequency component, presumably vortices. The CB concept is too simplified to describe such a two-component turbulence, and development of new approaches describing coexisting and interacting waves and vortices are needed. Interestingly, the CB prediction for the kinetic energy spectrum is observed for the wavenumbers both below and above $k_\perp=24$ without any particular feature observed at this transitional scale. One could argue that what is observed is just the classical Kolmogorov spectrum and not CB, but the evident strong anisotropy of the spectrum plays agains such a simple explanation.

\begin{figure}
  \centerline{\includegraphics[height=11cm, width=11cm]{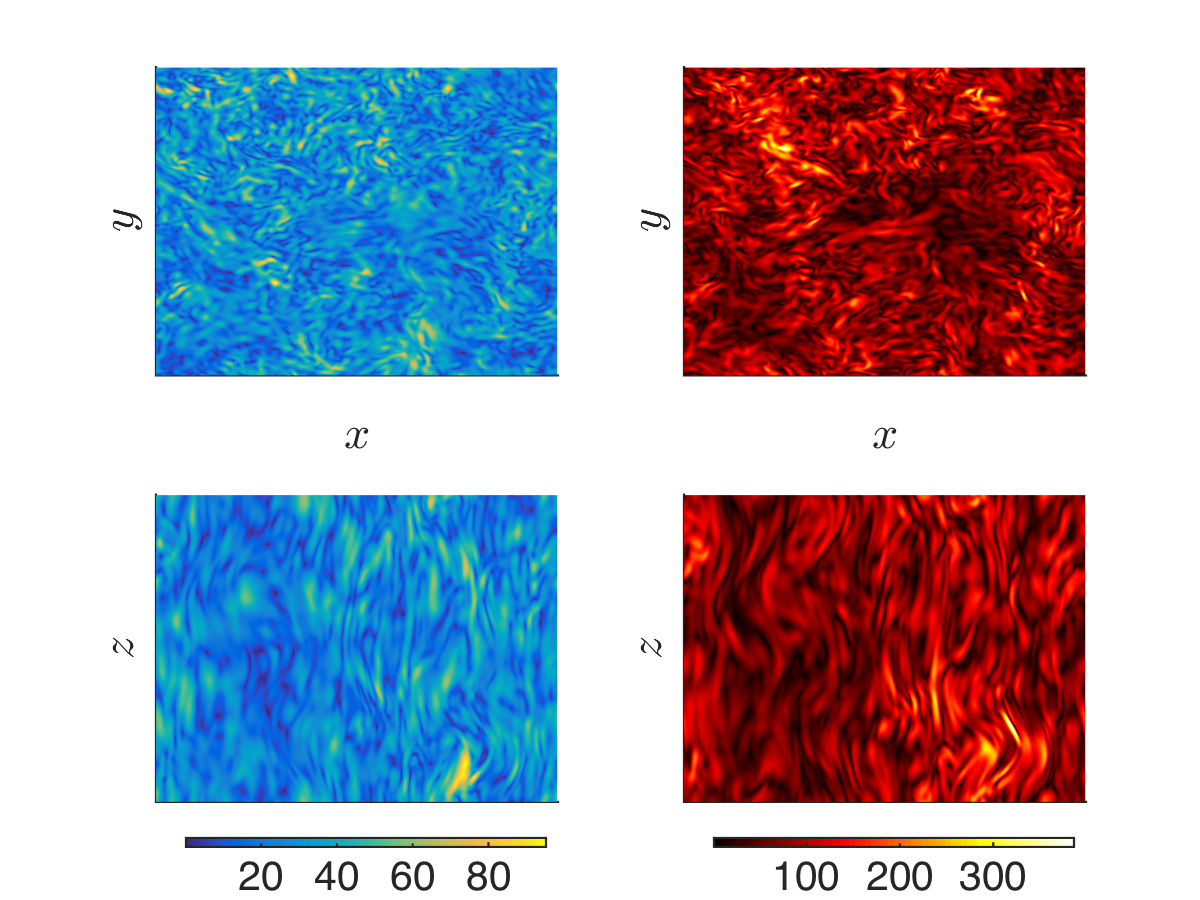}}
  \caption{ Simulation A fields in the physical space. \textit{Left:} vorticity amplitude in the 
  ${\bf x}_\perp = (x,y)$ plane and the $(x,z)$ plane.
   \textit{Right:} current amplitude in the   $\bf x_\perp$ plane and the $(x,z)$ plane.
}
\label{fig:CBvortcurr}
\end{figure}

In Figures~\ref{fig:CBvortcurr} we plot the amplitude of the vorticity and current both in the ${\bf x}_\perp =(x,y)$ plane and the $(x,z)$ plane. In the perpendicular plane, we see a collection of small-scale structures both in the current and in the vorticity which are isotropically and homogeneously distributed. In the vertical plane see that the structures are stretched in the parallel direction as expected for an anisotropic state. Combined with the $2D$ spectra in Figure~\ref{fig:CBAni}, we have strong evidence for the anisotropy assumption which both the weak and strong turbulence predictions are based upon. 

\subsection{Simulation $B$: Weak wave turbulence}

In simulation $B$ we have reduced amplitude, and magnetic and kinetic energy  (with respect  to simulation $A$) in an attempt to access a regime relevant to WWT. The full set of parameters are again given in table~\ref{tab:para}. With this parameter set, it was found that the development towards a stationary spectra was much slower than in simulation $A$. Due to this, the simulation has been performed with a resolution of $256^3$. 

\begin{figure}
  \centerline{\includegraphics[height=5cm, width=14cm]{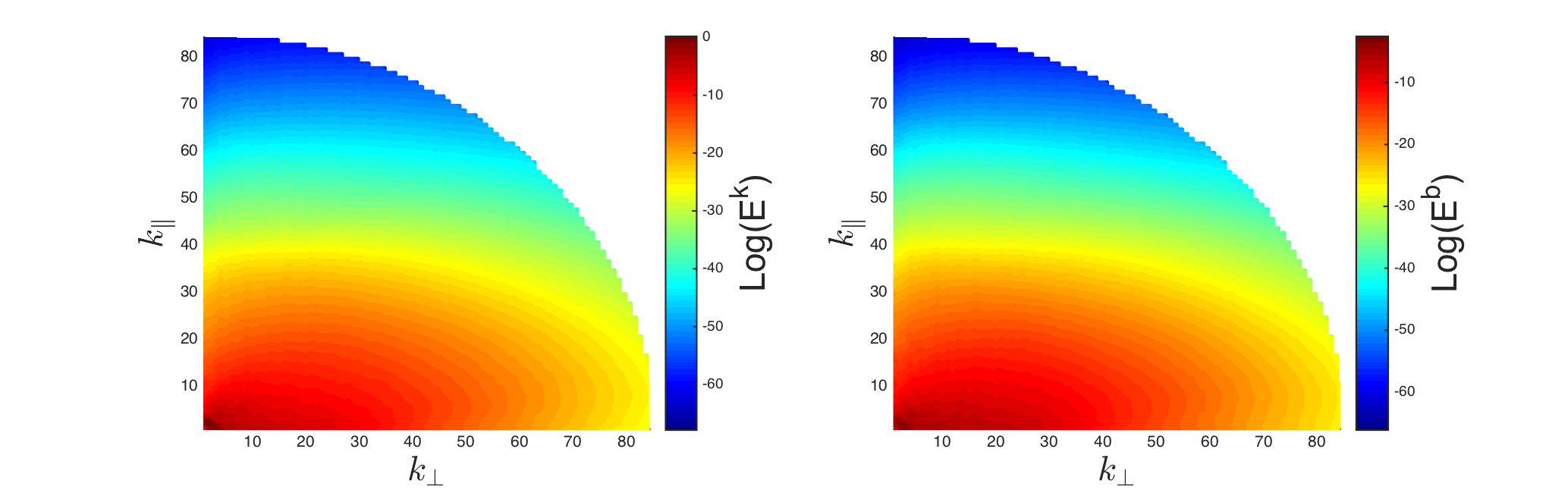}}
  \caption{$2D$ energy spectra for the (\textit{a}) kinetic\protect ~ and (\textit{b}) magnetic energies in simulation B.}
\label{fig:WTAni}
\end{figure}

\begin{figure}
  \centerline{\includegraphics[width=1.1\textwidth]{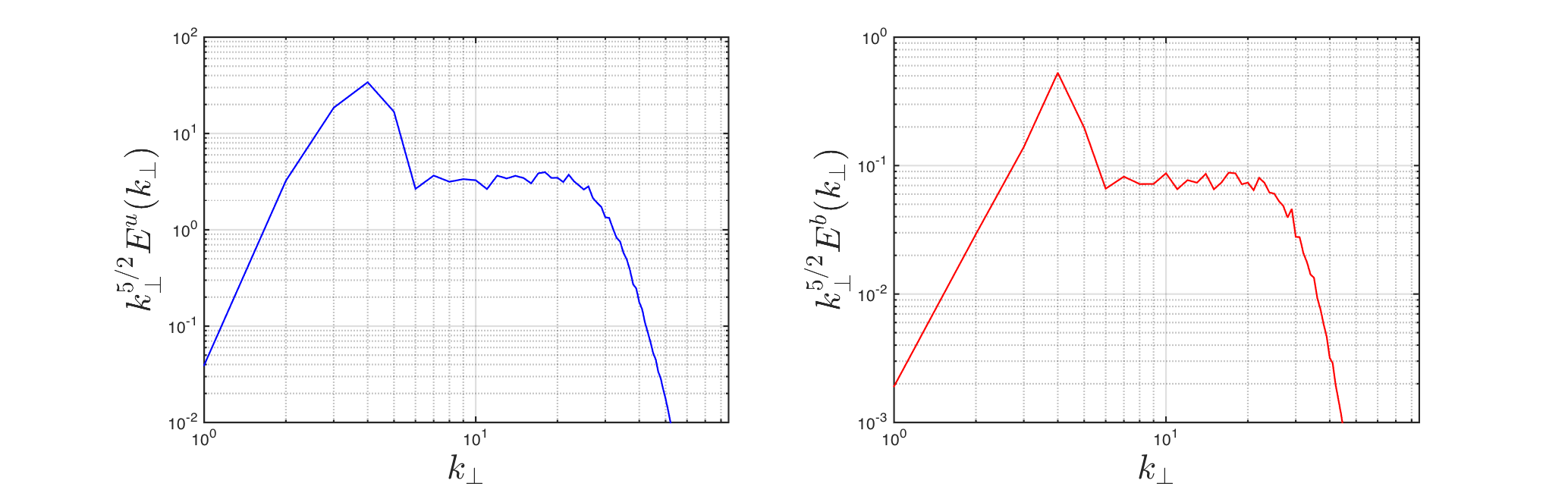}}
  \caption{Compensated axisymmetric energy spectra for the (\textit{a}) kinetic\protect ~ and (\textit{b}) magnetic energies in simulation B.}
\label{fig:WTaxispec}
\end{figure}

The $2D$ energy spectra for simulation $B$ are plotted in Figure~\ref{fig:WTAni}. As expected, for both the kinetic and magnetic energy there is a preferential transfer of energy along $k_\perp$. Figure~\ref{fig:WTaxispec} shows the axisymmetric energy spectra integrated over $k_\parallel$. Each spectrum has been compensated by the WWT prediction, $k_\perp^{-5/2}$,  and both appear to be flat, suggesting an agreement with the prediction.

\begin{figure}
  \centerline{\includegraphics[width=0.75\textwidth]{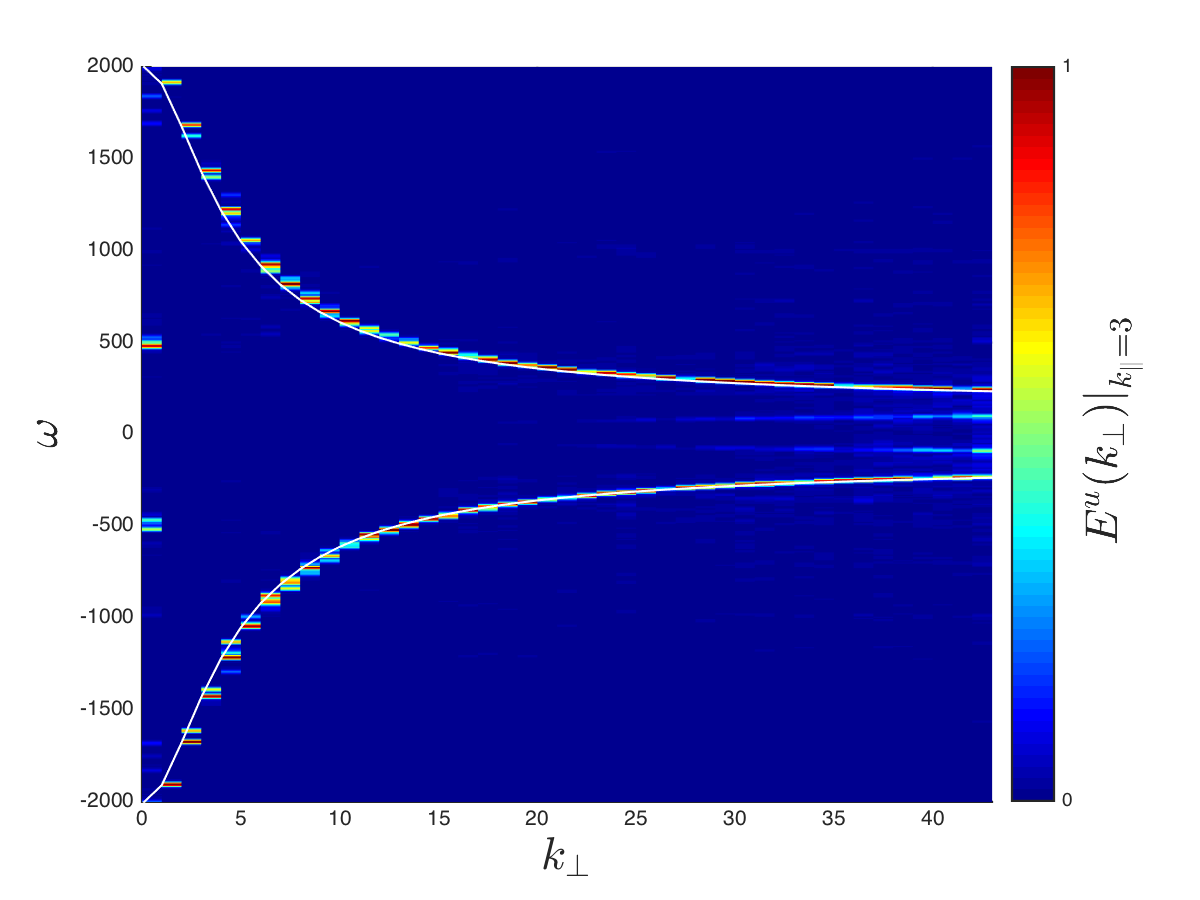}}
  \caption{Spatio-temporal energy kinetic energy spectrum in simulation B. The white lines indicate the dispersion relation for inertial waves.}
\label{fig:WTkinfreq}
\end{figure}

\begin{figure}
  \centerline{\includegraphics[width=0.75\textwidth]{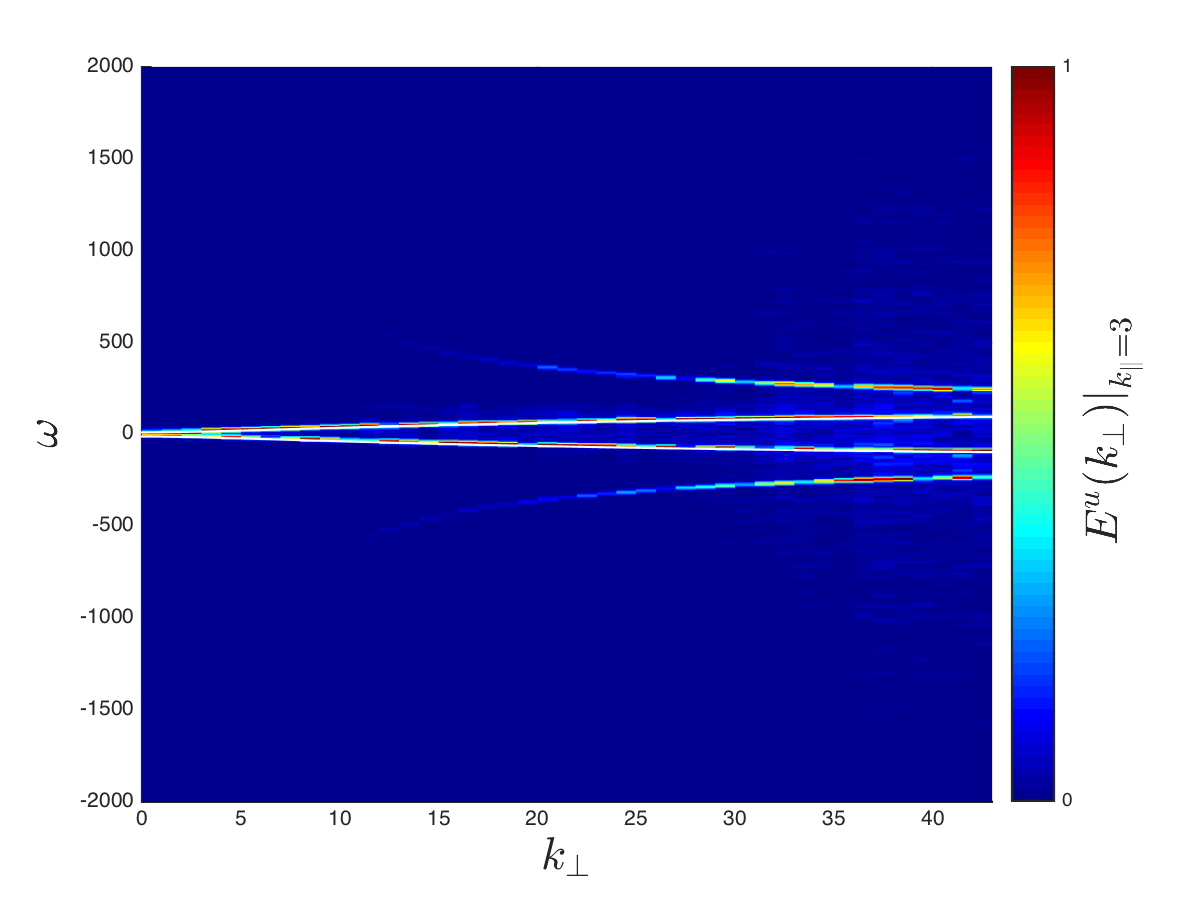}}
  \caption{Spatio-temporal energy magnetic energy spectrum in simulation B. The white lines indicate the dispersion relation for magnetostrophic waves.}
\label{fig:WTmagfreq}
\end{figure}

As before, the presence of waves can be tested by plotting the spatio-temporal spectra for the magnetic and kinetic energy. This is done in Figures~(\ref{fig:WTkinfreq}) and (\ref{fig:WTmagfreq}). They are plotted alongside the dispersion relation for inertial and megnetostrophic waves respectively. In each case, we see a build up of a narrow energy distribution on and near the dispersion relation suggesting the presence of weak inertial and megnetostrophic waves. This combined with the apparent agreement between the energy spectra and the WWT prediction, leads us to believe that the dynamics that the WWT regime is realised in simulation B. 

The physical space distributions of the vorticity and the  magnetic fields  in the simulation B appear to be similar (perhaps somewhat more anisotropic) as in Figure \ref{fig:CBvortcurr} for the simulation A; so we omit presenting separate images of these fields here.

\clearpage
\section{Conclusions}
In this paper we have studied turbulence in a rotating MHD system using direct numerical simulation of the governing equations. We have reviewed the  weak  wave turbulence in rotating MHD the theory of which was developed in~\citet{Galtier2014b}. We have also derived scalings for the kinetic and magnetic energy spectra  for strong wave turbulence based upon the CB phenomenology which postulates the timescales of the linear waves and the eddies are of comparable magnitude. 

Two simulations have been presented, one for a strong and a weak wave turbulence regimes respectively. As expected, and as previously observed for purely MHD and purely rotating turbulence cases, turbulence quickly evolves toward strongly anisotropic states with elongated along the external magnetic field vortex and current structures. In each case, we have presented the numerical results for the 2D (axial angle averaged) energy spectra, 1D energy spectra (obtained by integrating the 2D spectra over the parallel wavenumber) and spatio-temporal spectra at a fixed parallel wavenumber. For strong and weak wave turbulence, these results are in a good agreement with predictions of the CB and the WWT theoretical predictions, particularly in the part concerning the 1D spectra. The spatio-temporal plots have allowed us to differentiate between weak dispersive waves and strongly-turbulent structures, such as hydrodynamic vortices and, thereby, directly examine realisability of assumptions of weak nonlinearity and of CB. For instance, it allowed us the verify that the weak nonlinearity property in simulation B.  However, for the strong nonlinearity run (simulation A) the spatio-temporal spectrum has revealed that the CB picture, where the frequency broadening is of the same size as the linear frequency, is somewhat oversimplified. Namely, in the low-wavenumber range the the spatio-temporal spectrum has a well-pronounced two-peaked distribution with a narrow peak at the linear frequency and a wide peak around the zero frequency. This indicates that a finer picture of strong wave turbulence in this case should contain two distinct dynamical components, weak waves and strong vortices (a ``condensate") which coexist and interact. Developing such a description is an interesting subject for future research.

\section{Acknowledgements}
Nicholas Bell was  supported  by  EPSRC  grant  1499213  funding  the  PhD  study. Sergey Nazarenko is supported by the Chaire D’Excellence IDEX (Initiative of Excellence) awarded by Universit\'e de la C\^ote d’Azur, France.
\section{Appendix}

The  derivation of the WWT for rotating MHD  was presented in~\cite{Galtier2014b}, 
where the  wave kinetic equations (WKEs) were derived. This was followed by finding the KZ spectra for the inertial and magnetostrophic waves under an assumption that these two types of waves are decoupled from each other. However, the decoupling property was simply assumed. In this appendix we will explore the conditions under which the decoupling assumption is valid. Further, in the second part we will find a regime where the nonlinear coupling between the inertial and magnitostrophic waves is important and will present simplified WKEs for such a regime.

The full WKEs for for rotating MHD  are as follows~\cite{Galtier2014b},
\begin{align} \label{eq:fullkinrot}
\frac{\partial n_\Lambda^s(\vec{k})}{\partial t} &= \frac{\pi \epsilon^2 d^4}{64 b_0^2} \iint \sum_{\substack{\Lambda_1,\Lambda_2 \\ s_1,s_2}}\left( \frac{\sin \psi_k}{k}\right)^2 k^2 k_1^2 k_2^2 (\Lambda k +\Lambda_1 k_1 +\Lambda_2 k_2 )^2 \nonumber \\
& \quad \times (\xi_{\Lambda}^{s})^2 (\xi_{\Lambda_1}^{s_1})^2 (\xi_{\Lambda_2}^{s_2})^2 \left(\frac{\xi_{\Lambda_2}^{-s_2} - \xi_{\Lambda_1}^{-s_1}}{k_\parallel}\right)^2 \left( \frac{\omega_\Lambda^s}{1+(\xi_\Lambda^{-s})^2}\right) \nonumber \\
&\quad \times \left( 2+ (\xi_{\Lambda}^{-s})^2 (\xi_{\Lambda_1}^{-s_1})^2(\xi_{\Lambda_2}^{-s_2})^2 -(\xi_{\Lambda}^{-s})^2 - (\xi_{\Lambda_1}^{-s_1})^2 - (\xi_{\Lambda_2}^{-s_2})^2 \right)^2 \nonumber \\
& \quad \times \left[ \frac{\omega_{\Lambda}^{s}}{\lbrace 1+ (\xi_{\Lambda}^{-s})^2\rbrace n_{\Lambda}^{s}(\vec{k})} - \frac{\omega_{\Lambda_1}^{s_1}}{\lbrace 1+ (\xi_{\Lambda_1}^{-s_1})^2\rbrace n_{\Lambda_1}^{s_1}(\vec{k_1})} - \frac{\omega_{\Lambda_2}^{s_2}}{\lbrace 1+ (\xi_{\Lambda_2}^{-s_2})^2\rbrace n_{\Lambda_2}^{s_2}(\vec{k_2})}\right] \nonumber \\
&\quad\times n_{\Lambda}^{s}(\vec{k}) n_{\Lambda_1}^{s_1}(\vec{k_1}) n_{\Lambda_2}^{s_2}(\vec{k_2}) 
\delta(\omega_{\Lambda_1}^{s_1} + \omega_{\Lambda_2}^{s_2} -\omega_{\Lambda}^{s}
)\delta(\vec{k}-\vec{k_1}-\vec{k_2})\mathrm{d}\vec{k_1}\mathrm{d}\vec{k_2}.
\end{align}
The angle $\psi_k$ refers to the angle opposite $\vec{k}$ in the triangle defined by $\vec{k}=\vec{k_1}+\vec{k_2}$, $\xi_\Lambda^s$ is defined by
\begin{equation}
\xi_\Lambda^s = \frac{-skd}{-s\Lambda+ \sqrt{1+k^2 d^2}}.
\end{equation}
Here $s=\pm $ defines the directional wave polarity and $\Lambda=\pm s$ defines the circular polarization. If $\Lambda= s$ then we are dealing with the right polarized wave which is the magnetostrophic wave in this case. The left polarized waves are given by $\Lambda = -s$ and correspond to inertial waves. Solving the full kinetic equation would be extremely difficult even numerically. Thus, one tends to reduce the WKEs by considering by considering relevant special cases. 

First we shall consider the case when parameter $kd$ is small. 
Consider first the equation describing the dynamics of the inertial waves ($\Lambda=-s$). Performing the summation over polarisations $\Lambda_1$ and $\Lambda_2$, the kinetic equation takes the form
\begin{equation} \label{eq:kinexpinertial}
\partial_t n_{-s}^s(\vec{k})=A_{II}+B_{IM}+C_{MM},
\end{equation}
where $A_{II}$ gives the contribution from inertial-inertial wave interactions, $B_{IM}$ gives the contribution from inertial-magnetostrophic wave interactions and $C_{MM}$ gives the contribution from magnetostrophic-magnetostrophic interactions. Now, to leading order in $kd$ we have the following expansions, 
$
\xi_{-s}^s\rightarrow-\frac{skd}{2},
\xi_s^s\rightarrow-\frac{2s}{kd},
\omega_{s}^{s}\rightarrow \frac{s k_\parallel k d b_0}{2} =\omega_M,
\omega_{-s}^s\rightarrow \frac{2 s\Omega_0 k_\parallel}{k}=\omega_I.
$
Using these expansions, we write the asymptotic expressions for the terms in Equation~(\ref{eq:kinexpinertial}) as
\begin{align} \label{eq:AII}
A_{II}&=\frac{\pi\epsilon^2}{4b_0^2}\int\sum_{s_1,s_2}\left( \frac{\sin\psi_k}{k}\right)^2 \left(sk+s_1k_1+s_2k_2\right)^2\frac{\left(s_2k_1-s_1k_2\right)^2}{k_1^2k_2^2} \nonumber\\
&\qquad\times \frac{k^2\omega_{-s}^s}{k_\parallel^2}n_{-s}^s n_{-s_1}^{s_1} n_{-s_2}^{s_2} \left[ \frac{k^2\omega_{-s}^{s}}{n_{-s}^{s}}-\frac{k_1^2\omega_{-s_1}^{s_1}}{n_{-s_1}^{s_1}}-\frac{k_2^2\omega_{-s_2}^{s_2}}{n_{-s_2}^{s_2}} \right] \nonumber\\
&\qquad\times
\delta(\omega_{-s_1}^{s_1} + \omega_{-s_2}^{s_2} -\omega_{-s}^{s}
)\delta(\vec{k}-\vec{k_1}-\vec{k_2})\mathrm{d}\vec{k}_1\mathrm{d}\vec{k}_2,
\end{align}
\begin{align} \label{eq:BIM}
B_{IM}&=\frac{\pi\epsilon^2}{8 b_0^2}\int\sum_{s_1,s_1} k^2d^2 \left( \frac{\sin\psi_k}{k}\right)^2 \left(sk+s_1k_1+s_2k_2\right)^2\frac{(k_2^2-k_1^2-k^2)^2}{k_1^2} \frac{\omega_{-s}^s}{k_\parallel^2}n_{-s}^s n_{-s_1}^{s_1} n_{s_2}^{s_2} \nonumber\\
&\!\!\! \times\left[ \frac{k^2d^2\omega_{-s}^{s}}{4n_{-s}^{s}}-\frac{k_1^2d^2\omega_{-s_1}^{s_1}}{4n_{-s_1}^{s_1}}-\frac{\omega_{s_2}^{s_2}}{n_{s_2}^{s_2}} \right] \delta(\omega_{-s_1}^{s_1} + \omega_{s_2}^{s_2} -\omega_{-s}^{s}
)\delta(\vec{k}-\vec{k_1}-\vec{k_2})
\mathrm{d}\vec{k}_1\mathrm{d}\vec{k}_2,
\end{align}
\begin{align} \label{eq:CMM}
C_{MM}&=\frac{\pi\epsilon^2}{16b_0^2}\int\sum_{s_1,s_2}k^2d^2\left( \frac{\sin\psi_k}{k}\right)^2 \left(sk+s_1k_1+s_2k_2\right)^2\left(s_2k_2-s_1k_1\right)^2 \frac{\omega_{-s}^s}{k_\parallel^2}n_{-s}^s n_{s_1}^{s_1}n_{s_2}^{s_2} \nonumber \\
&\times\left[ \frac{k^2d^2\omega_{-s}^{s}}{4n_{-s}^{s}}-\frac{\omega_{s_1}^{s_1}}{n_{s_1}^{s_1}}-\frac{\omega_{s_2}^{s_2}}{n_{s_2}^{s_2}}\right] 
\delta(\omega_{s_1}^{s_1} + \omega_{s_2}^{s_2} -\omega_{-s}^{s}
)\delta(\vec{k}-\vec{k_1}-\vec{k_2})\mathrm{d}\vec{k}_1\mathrm{d}\vec{k}_2.
\end{align}

Now consider the equation describing the dynamics of the magnetostrophic wave action, $n_s^s$. The WKE will assume a similar form as for inertial waves,
\begin{equation} \label{eq:kinexpmagneto}
\partial_t n_s^s(\vec{k})=D_{MM}+E_{IM}+F_{II}.
\end{equation}
The above  expansions  are again used to give the individual terms as
\begin{align} \label{eq:DMM}
D_{MM}&=\frac{\pi\epsilon^2}{b_0^2}\int\sum_{s_1,s_2}\left( \frac{\sin\psi_k}{k}\right)^2 \left(sk+s_1k_1+s_2k_2\right)^2\left(s_1k_1-s_2k_2\right)^2\frac{\omega_s^s}{k_\parallel^2} \nonumber\\
&\times n_s^s n_{s_1}^{s_1} n_{s_2}^{s_2}\left[\frac{\omega_s^s}{n_s^s}-\frac{\omega_{s_1}^{s_1}}{n_{s_1}^{s_1}}-\frac{\omega_{s_2}^{s_2}}{n_{s_2}^{s_2}}\right]
\delta(\omega_{s_1}^{s_1} + \omega_{s_2}^{s_2} -\omega_{s}^{s}
)\delta(\vec{k}-\vec{k_1}-\vec{k_2})\mathrm{d}\vec{k}_1\mathrm{d}\vec{k}_2,
\end{align}
\begin{align} \label{eq:EIM}
E_{IM}&=\frac{8\pi\epsilon^2}{b_0^2}\int\sum_{s_1,s_2}\left( \frac{\sin\psi_k}{k}\right)^2 \left(sk+s_1k_1+s_2k_2\right)^2\frac{\omega_s^s}{k_\parallel^2 k_1^2 d^4}  n_s^s n_{-s_1}^{s_1} n_{s_2}^{s_2} \nonumber\\
&\times \left[\frac{\omega_s^s}{n_s^s}-\frac{k_1^2d^2\omega_{-s_1}^{s_1}}{4n_{-s_1}^{s_1}}-\frac{\omega_{s_2}^{s_2}}{n_{s_2}^{s_2}}\right]
\delta(\omega_{-s_1}^{s_1} + \omega_{s_2}^{s_2} -\omega_{s}^{s}
)\delta(\vec{k}-\vec{k_1}-\vec{k_2})\mathrm{d}\vec{k}_1\mathrm{d}\vec{k}_2,
\end{align}
\begin{align} \label{eq:FII}
F_{II}&=\frac{\pi\epsilon^2}{4b_0^2}\int\sum_{s_1,s_2}\left( \frac{\sin\psi_k}{k}\right)^2 \left(sk+s_1k_1+s_2k_2\right)^2\left(s_2k_1-s_1k_2\right)^2 \nonumber\\
&\times\frac{\left(k^2-k_1^2-k_2^2\right)^2}{k_1^2k_2^2}\frac{\omega_s^s}{k_\parallel^2}n_s^s n_{-s_1}^{s_1} n_{-s_2}^{s_2}\left[\frac{\omega_s^s}{n_s^s}-\frac{k_1^2d^2\omega_{-s_1}^{s_1}}{4n_{-s_1}^{s_1}}-\frac{k_2^2d^2\omega_{-s_2}^{s_2}}{4n_{-s_2}^{s_2}}\right] \nonumber\\
&\times
\delta(\omega_{-s_1}^{s_1} + \omega_{-s_2}^{s_2} -\omega_{s}^{s}
)\delta(\vec{k}-\vec{k_1}-\vec{k_2})\mathrm{d}\vec{k}_1\mathrm{d}\vec{k}_2.
\end{align}

\subsection{Decoupled Kinetic Equations} \label{DecoupledRMHD}
KZ spectra were derived in~\cite{Galtier2014b} assuming decoupling of the WKEs for the inertial and the magnetostrophic waves. Here, we will look at the decoupling property more closely. Let us assume that the dynamically important wavenumbers are of the same order for both the inertial waves and the magnetostrophic waves,
\begin{equation} \label{eq:scdec1}
k\sim k_\parallel \sim 1,
\end{equation}
the anisotropic assumption is made later. In terms of our small parameter $kd\equiv \lambda$, we get the following scalings, 
$
b_0\sim\lambda,
\Omega_0\sim 1,
\omega_s^s \sim \lambda^2,
\omega_{-s}^{s} \sim 1,
n_s^s\sim 1/\lambda^2,
n_{-s}^s \sim 1.
$
The scaling for the wave actions $n^s_s$ and $n_{-s}^s$ come from ensuring that the total energy contained within the inertial waves is of the same order as the total energy in the magnetostrophic waves,\begin{equation}
\int \omega_{-s}^s E_{-s}^s(\vec{k})\mathrm{d}\vec{k} \sim \int \omega_{s}^s E_{s}^s(\vec{k})\mathrm{d}\vec{k},
\end{equation}
together with the following relation for the energy given by~\cite{Galtier2014b}
\begin{equation}
E_\Lambda^s(\vec{k}) = \left[ 1 + (\xi_\Lambda^{-s})^2 \right]n_\Lambda^s(\vec{k}).
\end{equation}



Now one can use the above scalings to compare the relative magnitudes of the terms in Equation~(\ref{eq:kinexpinertial}). One can immediately neglect $C_{MM}$  because
$
\delta(\omega_{-s}^s-\omega_{s_1}^{s_1}-\omega_{s_2}^{s_2}) =0
$
if condition (\ref{eq:scdec1}) is satisfied (since in this case $\omega_{-s}^s \gg \omega_{s}^{s}$).
 The remaining two terms scale as 
$
A_{II}\sim 1/\lambda^2,
 B_{IM}\sim 1,
$
and thus $B_{IM}$ is negligible and we get the following WKE,
\begin{align} \label{InertialDecoupled}
\partial_t n_{-s}^s(\vec{k})&=\frac{\pi\epsilon^2}{4b_0^2}\int\sum_{s_1,s_2}\left( \frac{\sin\psi_k}{k}\right)^2 \left(sk+s_1k_1+s_2k_2\right)^2\frac{\left(s_2k_1-s_1k_2\right)^2}{k_1^2k_2^2} \nonumber \\
&\qquad\times \frac{k^2\omega_{-s}^s}{k_\parallel^2}n_{-s}^s n_{-s_1}^{s_1} n_{-s_2}^{s_2} \left[ \frac{k^2\omega_{-s}^{s}}{n_{-s}^{s}}-\frac{k_1^2\omega_{-s_1}^{s_1}}{n_{-s_1}^{s_1}}-\frac{k_2^2\omega_{-s_2}^{s_2}}{n_{-s_2}^{s_2}} \right] \nonumber \\
&\qquad\times
\delta(\omega_{-s_1}^{s_1} + \omega_{-s_2}^{s_2} -\omega_{s}^{s}
)\delta(\vec{k}-\vec{k_1}-\vec{k_2})\mathrm{d}\vec{k}_1\mathrm{d}\vec{k}_2.
\end{align}


Same reasoning can now be applied to Equation~(\ref{eq:kinexpmagneto}) describing the magnetostrophic waves. Therm $E_{IM}$ is zero, since
$
\delta(\omega_{s}^s-\omega_{-s_1}^{s_1}-\omega_{s_2}^{s_2})  =0,$
and 
$D_{MM}\sim1/\lambda^2,
F_{II}\sim 1.$
Thus, the term containing only magnetostrophic waves is dominant and the WKE becomes
\begin{align} \label{MagnetoDecoupled}
\partial_tn_s^s(\vec{k})&=\frac{\pi\epsilon^2}{b_0^2}\int\sum_{s_1,s_2}\left( \frac{\sin\psi_k}{k}\right)^2 \left(sk+s_1k_1+s_2k_2\right)^2\left(s_1k_1-s_2k_2\right)^2\frac{\omega_s^s}{k_\parallel^2} \nonumber\\
&\qquad\times n_s^s n_{s_1}^{s_1} n_{s_2}^{s_2}\left[\frac{\omega_s^s}{n_s^s}-\frac{\omega_{s_1}^{s_1}}{n_{s_1}^{s_1}}-\frac{\omega_{s_2}^{s_2}}{n_{s_2}^{s_2}}\right]
\delta(\omega_{s_1}^{s_1} + \omega_{s_2}^{s_2} -\omega_{s}^{s}
)\delta(\vec{k}-\vec{k_1}-\vec{k_2})\mathrm{d}\vec{k}_1\mathrm{d}\vec{k}_2.
\end{align}
Thus, the WKEs for the inertial and magnetostrophic waves are completely decoupled in the limit $kd\rightarrow 0$ if the wavenumbers of the both types of waves are of similar magnitudes and the energy densities are comparable. 

\subsection{Coupled Kinetic Equations}

A regime in which this decoupling did not occur such that there was a transfer of energy between the two types of waves could be interesting dynamically. This occurs when terms $C_{MM}$ and $E_{IM}$ are non-zero i.e. if $\omega_I=\omega_{-s}^s=\frac{2\Omega_0 s k_\parallel}{k}$ and $\omega_M=\omega_{s}^s=\frac{s k_\parallel k db_0}{2}$ are of the same order of magnitude. This can be achieved if we consider the wavenumbers for the inertial waves such that
$k=\mathcal{O}(1)\quad\mathrm{and}\quad k_\parallel=\mathcal{O}(\lambda^2).
$
whereas for the magnetostrophic wavenumbers we still have 
$k \sim k_\parallel=\mathcal{O}(1).
$
 The frequencies and the wave action spectra now scale as follows,
$
\omega_{-s}^{s}\sim \lambda^2,
\omega_{s}^{s}\sim \lambda^2,
n_{-s}^{s}\sim 1,
n_s^s\sim 1,
$
where once again the energy in the inertial and magnetostrophic waves is assumed to be of the same order of magnitude. 

Now that the frequencies are all of the same order, no terms are zero due to the frequency resonance condition. However, $B_{IM}$ and $F_{II}$ are zero due to the   delta-function
$ \delta(\vec{k}-\vec{k_1}-\vec{k_2}) 
= \delta(\vec{k_\perp}-\vec{k}_{1\perp}-\vec{k}_{2\perp})\delta(k_\parallel - k_{1\parallel} - k_{2\parallel}).
$
The parallel wavenumber delta-function in $B_{IM}$ is zero because  $k_{2\parallel}$ is large whereas $k_{\parallel}$ and $k_{1\parallel}$ are small. The $F_{II}$ term is zero due to the same argument. The remaining terms scale as $ A_{II}\sim 1/\lambda^2,
C_{MM}\sim 1/\lambda^2,
D_{MM}\sim 1,
E_{IM}\sim 1/\lambda^2.
$
From this we have to leading order,
\begin{align} \label{CoupledInertial}
\partial_t n_{-s}^s(\vec{k})&=\frac{\pi\epsilon^2}{4b_0^2}\int\sum_{s_1,s_2}\left( \frac{\sin\psi_k}{k}\right)^2 \left(sk+s_1k_1+s_2k_2\right)^2\frac{\left(s_2k_1-s_1k_2\right)^2}{k_1^2k_2^2} \nonumber\\
&\qquad\times \frac{k^2\omega_{-s}^s}{k_\parallel^2}n_{-s}^s n_{-s_1}^{s_1} n_{-s_2}^{s_2} \left[ \frac{k^2\omega_{-s}^{s}}{n_{-s}^{s}}-\frac{k_1^2\omega_{-s_1}^{s_1}}{n_{-s_1}^{s_1}}-\frac{k_2^2\omega_{-s_2}^{s_2}}{n_{-s_2}^{s_2}} \right] \nonumber\\
&\qquad\times
\delta(\omega_{-s_1}^{s_1} + \omega_{-s_2}^{s_2} -\omega_{-s}^{s}
)\delta(\vec{k}-\vec{k_1}-\vec{k_2})\mathrm{d}\vec{k}_1\mathrm{d}\vec{k}_2 \nonumber\\
&-\frac{\pi\epsilon^2}{16b_0^2}\int\sum_{s_1,s_2}k^2d^2\left( \frac{\sin\psi_k}{k}\right)^2 \left(sk+s_1k_1+s_2k_2\right)^2\left(s_2k_2-s_1k_1\right)^2 \frac{\omega_{-s}^s}{k_\parallel^2}\nonumber \\
&\times n_{-s}^s n_{s_1}^{s_1}n_{s_2}^{s_2}\left[ \frac{\omega_{s_1}^{s_1}}{n_{s_1}^{s_1}}+\frac{\omega_{s_2}^{s_2}}{n_{s_2}^{s_2}}\right] 
\delta(\omega_{s_1}^{s_1} + \omega_{s_2}^{s_2} -\omega_{-s}^{s}
)\delta(\vec{k}-\vec{k_1}-\vec{k_2})\mathrm{d}\vec{k}_1\mathrm{d}\vec{k}_2.
\end{align}
Similarly, the  magnetostrophic WKE is
\begin{align} \label{CoupledMagneto}
\partial_t n_s^s(\vec{k})&=\frac{8\pi\epsilon^2\Omega^4}{b_0^6}\int\sum_{s_1,s_2}\left( \frac{\sin\psi_k}{k}\right)^2 \left(sk+s_1k_1+s_2k_2\right)^2\frac{\omega_s^s}{k_\parallel^2 k_1^2 } \nonumber\\
&\times n_s^s n_{-s_1}^{s_1} n_{s_2}^{s_2}\left[\frac{\omega_s^s}{n_s^s}-\frac{\omega_{s_2}^{s_2}}{n_{s_2}^{s_2}}\right]
\delta(\omega_{-s_1}^{s_1} + \omega_{s_2}^{s_2} -\omega_{s}^{s}
)\delta(\vec{k}-\vec{k_1}-\vec{k_2})\mathrm{d}\vec{k}_1\mathrm{d}\vec{k}_2.
\end{align}
Equations~(\ref{CoupledInertial}) and~(\ref{CoupledMagneto})  describe the kinetics of the inertial and magnetostrophic wave action spectra in a regime where there is coupling between the two types of wave. This regime is realised when there is a scale separation between the waves such that the perpendicular wavenumber dominates the parallel wavenumber in inertial range but are of the same scale in magnetostrophic waves. This setup points at an interesting dynamical regime where there may be significant transfer between kinetic and magnetic energy.

\bibliographystyle{plainnat}

\bibliography{Bibliography}

\begin{thebibliography}{32}
\providecommand{\natexlab}[1]{#1}
\providecommand{\url}[1]{\texttt{#1}}
\expandafter\ifx\csname urlstyle\endcsname\relax
  \providecommand{\doi}[1]{doi: #1}\else
  \providecommand{\doi}{doi: \begingroup \urlstyle{rm}\Url}\fi

\bibitem[Bou\'e et~al.(2011)Bou\'e, Dasgupta, Laurie, L'vov, Nazarenko, and
  Procaccia]{PhysRevB.84.064516}
Laurent Bou\'e, Ratul Dasgupta, Jason Laurie, Victor L'vov, Sergey Nazarenko,
  and Itamar Procaccia.
\newblock Exact solution for the energy spectrum of kelvin-wave turbulence in
  superfluids.
\newblock \emph{Phys. Rev. B}, 84:\penalty0 064516, Aug 2011.
\newblock \doi{10.1103/PhysRevB.84.064516}.
\newblock URL \url{https://link.aps.org/doi/10.1103/PhysRevB.84.064516}.

\bibitem[Finlay(2008)]{Finlay2008}
C.~C. Finlay.
\newblock Waves in the presence of magnetic fields, rotation and convection.
\newblock In P.~Cardin, editor, \emph{Dynamos}, volume~88, pages 403--450.
  Elsevier science publishers, 2008.

\bibitem[Galtier(2003)]{Galtier2003}
S.~Galtier.
\newblock Weak inertial-wave turbulence theory.
\newblock \emph{Phys. Rev. E}, 68\penalty0 (015301), 2003.

\bibitem[Galtier(2006)]{Galtier2006}
S.~Galtier.
\newblock Wave turbulence in incompressible hall magnetohydrodynamics.
\newblock \emph{J. Plasma Phys.}, 72\penalty0 (5):\penalty0 721--769, October
  2006.

\bibitem[Galtier(2014)]{Galtier2014b}
S.~Galtier.
\newblock Weak turbulence theory for rotating magnetohydrodynamics and
  planetary flows.
\newblock \emph{J. Fluid Mech.}, 757:\penalty0 114--154, 2014.

\bibitem[Galtier and Bhattacharjee(2005)]{Galtier2005}
S.~Galtier and A~Bhattacharjee.
\newblock Anisotropic wave turbulence in electron mhd.
\newblock \emph{Plasma Phys. Control Fusion}, \penalty0 (47):\penalty0
  B791--B701, 2005.

\bibitem[Galtier et~al.(2000)Galtier, Nazarenko, Newell, and
  Pouquet]{Galtier2000}
S.~Galtier, S.~Nazarenko, A.~C. Newell, and A.~Pouquet.
\newblock A weak turbulence theory for incompressible mhd.
\newblock \emph{J. Plasma Phys.}, 63\penalty0 (447), 2000.

\bibitem[Galtier et~al.(2001)Galtier, Nazarenko, and Newell]{Galtier2001}
S.~Galtier, S.~Nazarenko, and A.~C. Newell.
\newblock On wave turbulence in mhd.
\newblock \emph{Nonlinear Processes in Geophysics}, 8\penalty0 (3):\penalty0
  141--150, May 2001.

\bibitem[Goldreich and Sridhar(1995)]{Goldreich1995b}
P.~Goldreich and S.~Sridhar.
\newblock Toward a theory of interstellar turbulence. 2: strong alfv\'{e}nic
  turbulence.
\newblock \emph{Astrophys. J.}, 438\penalty0 (2):\penalty0 763--775, 1995.

\bibitem[G\'{o}mez et~al.(2005)G\'{o}mez, Mininni, and Dmitruk]{Gomez2005}
D.~O. G\'{o}mez, P.~D. Mininni, and P.~Dmitruk.
\newblock Parallel simulations in turbulent mhd.
\newblock \emph{Physica Scripta}, \penalty0 (T116):\penalty0 123, 2005.

\bibitem[Iroshnikov(1964)]{Iroshnikov1964}
P.~S. Iroshnikov.
\newblock Turbulence of a conducting fluid in a strong magnetic field.
\newblock \emph{Sov. Astron.}, 7:\penalty0 566--571, 1964.

\bibitem[Kolmogorov(1941)]{Kolmogorov1941a}
A.~N. Kolmogorov.
\newblock The local structure of turbulence in incompressible viscous fluid for
  very large reynolds numbers.
\newblock \emph{Dokl. Akad. Nauk. SSSR}, 30:\penalty0 301--305, 1941.

\bibitem[Kraichnan(1965)]{Kraichnan1965}
R.~H. Kraichnan.
\newblock Inertial range spectrum in hydromagnetic turbulence.
\newblock \emph{Phys. Fluids}, 8:\penalty0 1385--1387, 1965.

\bibitem[Leoni et~al.(2015)Leoni, Cobelli, and Mininni]{Leoni2015}
P.~C. Leoni, P.~J. Cobelli, and P.~D. Mininni.
\newblock The spatio-temporal spectrum of turbulent flows.
\newblock \emph{The European Physical Journal E}, 38\penalty0 (136), 2015.

\bibitem[Meyrand et~al.(2015)Meyrand, Kiyani, and Galtier]{Meyrand2015}
R.~Meyrand, K.~Kiyani, and S.~Galtier.
\newblock Intermittency in weak mhd turbulence.
\newblock \emph{J. Fluid Mech.}, 770, 2015.

\bibitem[Meyrand et~al.(2016)Meyrand, Galtier, and Kiyani]{Meyrand2016}
R.~Meyrand, S.~Galtier, and K.~Kiyani.
\newblock Direct evidence of the transition from weak to strong
  magnetohydrodynamic turbulence.
\newblock \emph{Phys. Rev. Lett.}, 116\penalty0 (105002), 2016.

\bibitem[Meyrand et~al.(2017)Meyrand, Kiyani, Gurcan, and Galtier]{Meyrand2017}
R.~Meyrand, K.~Kiyani, O.~D. Gurcan, and S.~Galtier.
\newblock Coexistence of weak and strong wave turbulence in incompressible hall
  magnetohydrodynamics.
\newblock \emph{arXiv:1712.10002}, 2017.

\bibitem[Mininni and Pouquet(2007)]{Mininni2007}
P.~D. Mininni and A.~Pouquet.
\newblock Energy spectra stemming from interactions of alfv{\'e}n waves and
  turbulent eddies.
\newblock \emph{Phys. Rev. Lett.}, 99\penalty0 (254502), 2007.

\bibitem[Mininni and Pouquet(2010)]{Mininni2010a}
P.~D. Mininni and A.~Pouquet.
\newblock Rotating helical turbulence. i. global evolution and spectral
  behaviour.
\newblock \emph{Phys. Fluids}, 22 (3)\penalty0 (035105), 2010.

\bibitem[Mininni et~al.(2011)Mininni, Rosenberg, Reddy, and
  Pouquet]{Mininni2011}
P.~D. Mininni, D.~Rosenberg, R.~Reddy, and A.~Pouquet.
\newblock A hybrid mpi--openmp scheme for scalable parallel pseudospectral
  computations for fluid turbulence.
\newblock \emph{Parallel Computing}, 37\penalty0 (6):\penalty0 316--326, 2011.

\bibitem[Mininni et~al.(2012)Mininni, Rosenberg, and Pouquet]{Mininni2012}
P.~D. Mininni, D.~Rosenberg, and A.~Pouquet.
\newblock Isotropization at small scales of rotating helically driven
  turbulence.
\newblock \emph{J. Fluid Mech.}, 699:\penalty0 263--279, 2012.

\bibitem[Nazarenko(2007)]{Nazarenko2007}
S.~Nazarenko.
\newblock 2d enslaving of mhd turbulence.
\newblock \emph{New J. Phys.}, 9\penalty0 (307), 2007.

\bibitem[Nazarenko(2011)]{NazarenkoBook}
S.~Nazarenko.
\newblock \emph{Wave Turbulence}.
\newblock Lecture Notes in Physics. Springer, 2011.

\bibitem[Nazarenko and Schekochichin(2011)]{Nazarenko2011}
S.~Nazarenko and A.~Schekochichin.
\newblock Critical balance in magnetohydrodynamic, rotating and stratified
  turbulence: towards a universal scaling conjecture.
\newblock \emph{J. Fluid Mech.}, 677:\penalty0 134, 2011.

\bibitem[Nazarenko and Onorato(2006)]{NAZARENKO20061}
Sergey Nazarenko and Miguel Onorato.
\newblock Wave turbulence and vortices in bose–einstein condensation.
\newblock \emph{Physica D: Nonlinear Phenomena}, 219\penalty0 (1):\penalty0 1
  -- 12, 2006.
\newblock ISSN 0167-2789.
\newblock \doi{https://doi.org/10.1016/j.physd.2006.05.007}.
\newblock URL
  \url{http://www.sciencedirect.com/science/article/pii/S0167278906001758}.

\bibitem[Nazarenko and Onorato(2007)]{Nazarenko2007a}
Sergey Nazarenko and Miguel Onorato.
\newblock Freely decaying turbulence and Bose--Einstein condensation in
  Gross--Pitaevski model.
\newblock \emph{Journal of Low Temperature Physics}, 146\penalty0 (1):\penalty0
  31--46, 2007.

\bibitem[Roberts and Belmont(2013)]{Roberts2013}
P.~H. Roberts and G.~Belmont.
\newblock On the genesis of the earth's magnetism.
\newblock \emph{Rep. Prog. Phys.}, 76 (9), 2013.

\bibitem[Schekochichin and Nazarenko(2012)]{Schekochihin2012}
A.~Schekochichin and S.~Nazarenko.
\newblock Weak alfv\'{e}n wave turbulence revisited.
\newblock \emph{Phys. Rev. E}, 85\penalty0 (3), March 2012.

\bibitem[Shirley and Fairbridge(1997)]{Shirley1997}
J.~H. Shirley and R.~W. Fairbridge.
\newblock \emph{Encyclopedia of Planetary Sciences}.
\newblock Springer, 1997.

\bibitem[Teitelbaum and Mininni(2012)]{Teitelbaum2009}
T.~Teitelbaum and P.~D. Mininni.
\newblock Effect of helicity and rotation on the free decay of turbulent flows.
\newblock \emph{Phys. Rev. Lett.}, 103 (1)\penalty0 (014501), 2012.

\bibitem[TenBarge and Howes(2011)]{TenBarge2012}
J.~M. TenBarge and G.~G. Howes.
\newblock Evidence of critical balance in kinetic alfv\'{e}n wave turbulence
  simulations.
\newblock \emph{Physics of Plasmas}, 19\penalty0 (5), 2011.

\bibitem[Zakharov et~al.(1992)Zakharov, L'vov, and G.]{ZakharovBook}
V.~E. Zakharov, V.~S. L'vov, and Falkovich. G.
\newblock \emph{Kolmogorov Spectra of Turbulence 1: Wave Turbulence}.
\newblock Springer, Berlin, 1992.

\end{thebibliography}

\end{document}